\documentclass[proceedings, preprint]{rmaa}

\SetYear{\today}
\SetConfTitle{Prospects for low-frequency radio astronomy in S. A.}

\title{End to end developments for the Multipurpose Interferometer Array Pathfinder from the IAR Electronics Laboratory} 

\author{
  J. M. Gonzalez,\altaffilmark{1} 
  H. Command,\altaffilmark{1}
  and G. Valdez\altaffilmark{1}}

\altaffiltext{1}{Instituto Argentino de Radioastronomía,
  IAR, Camino Gral. Belgrano Km 40, Berazategui,
  Buenos Aires, Argentina.}

\shortauthor{Gonzalez, Command, \& Valdéz}
\shorttitle{End to end developments for the MIA Pathfinder}

\listofauthors{J. M. Gonzalez, H. Command, \& G. Valdez}
\indexauthor{Gonzalez, J. M.}
\indexauthor{Command, H.}
\indexauthor{Valdez, G.}

\abstract{
The Multipurpose Interferometer Array Pathfinder (MIA), developed from the Argentine Institute of Radio Astronomy (IAR), is a radio astronomical instrument based on interferometry techniques, designed for the detection of radio emission from astronomical sources. Phase one consists of 16 antennas of 5 meters in diameter, with the possibility of increasing their number. In addition, it is equipped with a dual polarization receiver with a bandwidth of 250 MHz, centered at 1325 MHz, and a digitizer and processor for the correlation functions.
For the development of this instrument, a three antenna pathfinder is currently being built with its positioning control, radio frequency systems, acquisition and processing stages.
This paper will describe the concept design and their current progress for each stage.
}

\resumen{
El Multipurpose Interferometer Array Pathfinder (MIA), desarrollado desde el Instituto Argentino de Radioastronomía (IAR), es un instrumento radioastronómico basado en técnicas de interferometría, diseñado para la detección de emisión en radio de fuentes astronómicas. La primera fase consta de 16 antenas de 5 metros de diámetro, con posibilidad de aumentar su número. Además, está equipado con un receptor de polarización dual con un ancho de banda de 250 MHz, centrado en 1325 MHz, y un digitalizador y procesador para las funciones de correlación.
Para el desarrollo de este instrumento se está construyendo actualmente un pathfinder de tres antenas con sus etapas de control de posicionamiento, sistemas de radiofrecuencia, adquisición y procesado.
En este trabajo se describirá el diseño conceptual y sus avances actuales para cada etapa.

}

\addkeyword{Interferometer Array}
\addkeyword{Radioastronomy}
\addkeyword{Instrumentation}

\begin{document}
\maketitle

\section{Introduction}



Since the detection of radio signals from astronomical sources, radio astronomy has taken on a remarkable importance in the scientific community \citep{rohlfs2013tools,martin1983assessing}.

On the other hand, the advantages of interferometric techniques over the detection of such signals by single-dish antenna methods have led to the development of multiple interferometers in recent years.
\citep{kocz2019dsa,foley2016engineering,Hickish2016Adecade}. Because of this, the Instituto Argentino de Radioastronomía plans to develop the Multipurpose Interferometer Array (MIA), located in Argentina.
Due to the characteristics of the design and its location, it is estimated that this instrument will make a major contribution to the scientific community.

This paper details the status of the MIA design. The following section details the development of the Antenna and Front-End. Then it explains the status of the Back-end stage. Section 4 details the control algorithm used for the positioning of the antenna. Finally, the conclusions obtained are developed.

\section{Antenna and Front-End}

The MIA dishes, designed with a prime-focus alt-az configuration and a focal
ratio ($\frac {f}{D}$) of 0.43, are equipped with a dual polarization receiver centered at 1.5 GHz. This is complemented by a dual polarization feed operating at the same frequency. The feed's illumination efficiency is characterized by a '10 dB edge taper', ensuring an optimal balance between spillover and sub-illumination. In this context, 'f' represents the focal length of the paraboloid, which is the distance from the vertex to the focus, and 'D' denotes its diameter, the maximum distance across the reflector. To achieved the best cost-benefit ratio, 5 m of the antenna diameter was chosen \citep{dreher2000the}.

\begin{figure}
	\centering
	\includegraphics[width=1\columnwidth]{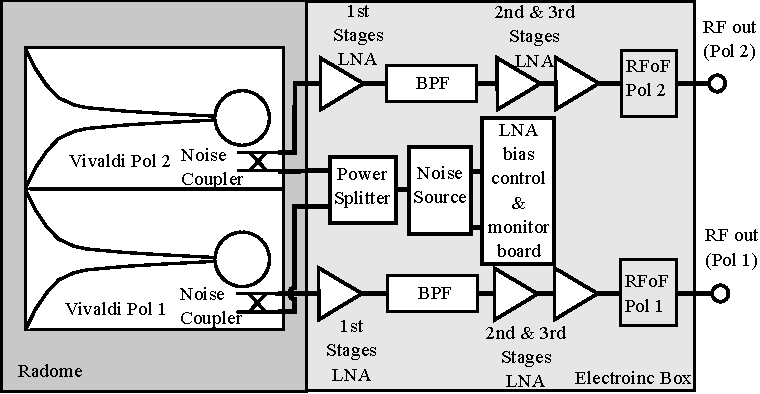}
	\caption{Front-End System Block Diagram.}
	\label{fig.front-end}
\end{figure}

The Front-End component of MIA is depicted in Figure~\ref{fig.front-end}. It is composed of two Vivaldi antennas, set at 90 degrees to each other, each corresponding to a distinct polarization. These antennas are equipped with a directional coupler, into which the required noise level is introduced for calibration purposes. The output of each antenna is connected to a primary stage of low-noise amplifiers (LNA), followed by a bandpass filter with a bandwidth ranging from 1000 MHz to 2000 MHz. Subsequently, the signal undergoes a secondary amplification stage utilizing LNA. The receiver features an equivalent noise temperature of approximately 50 K and a gain of 60 dB. The bandwidth is determined by the chosen bandpass filter. The receiver enclosure will be at ambient temperature with temperature control implemented via Peltier cells on critical components, such as the first-stage amplifiers and the noise-generating diode for calibration. RF output will be transmitted via optical fiber.

Vivaldi antennas are an attractive choice for dual-polarization radio interferometry applications due to their ultra-wideband capabilities, uniform radiation patterns, and ease of implementing dual polarization. Their compact and modular design simplifies fabrication and assembly while maintaining good performance across a wide frequency range. Despite some limitations, such as performance at lower frequencies and sensitivity to substrate properties, careful design and substrate selection can mitigate these issues, making Vivaldi antennas a suitable option for radio astronomy applications.

\begin{figure}
	\centering
	\includegraphics[width=0.8\columnwidth]{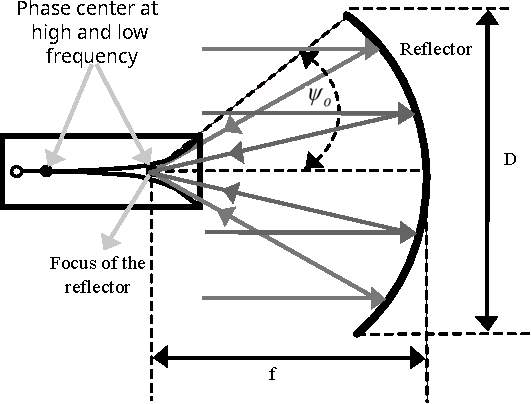}
	\caption{Vivaldi Reflector.}
	\label{fig.vivaldi_reflector}
\end{figure}

Figure~\ref{fig.vivaldi_reflector} detailed the Vivaldi reflector scheme. There, phase center location changes with changes in the frequency and in any wide band application, phase error losses are unavoidable. The left dot on the Vivaldi antenna corresponds to the phase center position for the highest frequencies, while the right dot for the lowest, which also aligns with the parabolic reflector's focal point. With its current positioning, it would be possible to use the reflector system perfectly at low frequencies of the band. However, at high frequencies there will be distance between the phase center location and the focus point of the reflector. This will result in phase error losses in the system. The half-subtended angle of the reflector $\psi_{0}$ is related with $\frac{f}{D}$ by,

\begin{equation}
        \frac{f}{D}=\frac{1}{4}cot\frac{\psi_{0}}{2}.
\end{equation}
 


Unequal phase center locations in E and H-planes introduce phase error losses due to astigmatism. It is detected by the depth of the nulls in the E and H-planes. Phase error loss due to astigmatism is not as severe as the losses due to axial defocusing.

\begin{figure}
	\centering
	\includegraphics[width=0.9\columnwidth]{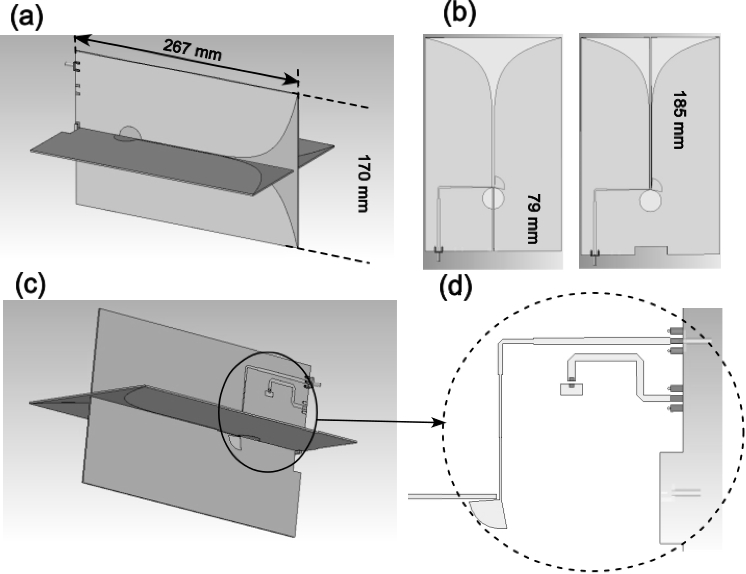}
	\caption{Vivaldi Assembly. (a) Optimized dimensions of two orthogonal Vivaldi antennas in a cross-shaped form. (b) Longitudinal cuts are made to make the insertion of both antennas possible. (c) and (d) Detailed view of three-stage quarter-wave balun feed and noise injection RF coupler.}
	\label{fig.vivaldi_assembly}
\end{figure}

Vivaldi is an end fire radiator supported on a thin substrate with a relative permittivity ($\epsilon_r$) of 4.5. This substrate is made of FR4, a glass-reinforced epoxy laminate material, with a thickness of t = 1.53 mm. Despite the completely planar geometry of Vivaldi, it can produce almost symmetric radiation patterns in the E and H-planes. The antenna is designed to operate in the band of 1 GHz – 2.3 GHz. A lower cut-off wavelength can be defined where the maximum separation of the radiators at the distant end is a half-wavelength.
Constant beamwidth requires an aerial to have a shape that can be completely specified in terms of dimensionless normalized wavelength units.
The dual polarization is achieved by orthogonally placing two Vivaldi antennas along their outer edges as shown in Figure~\ref{fig.vivaldi_assembly} (a). To facilitate the insertion of both antennas, longitudinal cuts are introduced. Specifically, one cut measures 185 mm and is positioned on the front side of one of the Vivaldi antennas, while the other cut, measuring 79 mm, is located on the rear side of the other Vivaldi. These specific dimensions and placements ensure that the antennas interlock appropriately, thus forming the dual Vivaldi antenna configuration, as detailed in Figure~\ref{fig.vivaldi_assembly} (b). The antenna feeding consists of a microstrip-to-slot line balun for wideband matching as depicted in Figure~\ref{fig.vivaldi_assembly} (c) and (d). A four-stage quarter-wave balun is designed to transform the 50 $\Omega$ microstrip line feeding to the 120-140 $\Omega$ slot line impedance.

\section{Back-end}

\begin{figure}
		\centering
		\includegraphics[width=0.9\columnwidth]{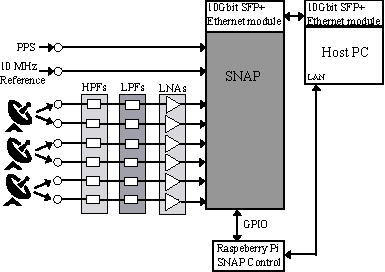}
		\caption{Back-End diagram.}
		\label{fig.Backend}
	\end{figure}

The analogue signals presented are digitized and packetized in the Back-End module using the Smart Network ADC Processor (SNAP\footnote{https://casper.berkeley.edu/wiki/SNAP}) boards, as is shown in Figure~\ref{fig.Backend}. The SNAP board is been conformed by three HMCAD1511 8-bit analog to digital converters (ADCs), capable of sampling two inputs each at 500 MHz for a total of 6 signals per board. The ADCs are connected to a Kintex-7 160T field-programmable gate array (FPGA), with an associated dual 10 GbE port. The SNAP board is controlled via Raspberry Pi, which interacts with a PC over ethernet via PYTHON scripts. The firmware for the hardware interfaces for the SNAP board is provided by the updated JASPER fork of the CASPER toolflow  \citep{parsons2008scalable}. 

In Figure~\ref{fig.Backend} the PPS signal represents the synchronism signal of the system. 
The 10 MHz signal is used for ADC calibration.
For the initial stage, it is planned to process the signal coming from 3 antennas, for which two polarizations are obtained, using 6 ADC inputs of the SNAP board. 
Between the output of the antennas and the inputs of the SNAP board, bandpass filters are placed, consisting of a low-pass filter and a high-pass filter, and also a low noise amplifier. 

On the other hand, the board can transmit data through a 1Gb/s and a 10 Gb/s ethernet ports.

\begin{figure}
		\centering
		\includegraphics[width=1\columnwidth,keepaspectratio]{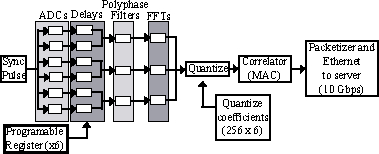}
		\caption{Correlator system diagram.}
		\label{fig.correlator}
	\end{figure}

Figure~\ref{fig.correlator} shows the processing stages in the FPGA. There, the three pairs of polarisations are sampled by 6 ADCs. For this, the ADCs operate at a frequency of 500 MHz, achieving an effective bandwidth of 250 MHz, at a centre frequency of 1325 MHz. These sampled signal pairs are then synchronised to maintain signal coherence. 

Once the signals are aligned, the CASPER libraries are used to transform them to the frequency domain. For this, a polyphase bank filter (PBA), implemented through a finite response filter (FIR), and the fast Fourier transform (FFT) are used. 
The FIR has a 512-point four-tap Hamming window response (for 256 spectral channels). The filter coefficients are 18 bits, thus allowing the 8-bit ADC data to be increased to a maximum of 18 real bits and 18 imaginary bits.

Once the signal is transformed to the frequency domain, it is quantified in 6 channels of 256 points. The correlation process is carried out in parallel by the multiply and accumulate (MAC) block belonging to the CASPER group libraries. 
Then, the generated signal passes through a packetizer and a first in-first out (FIFO) block, where the data packet is conditioned. At least, the data is transmitted at 10 Gb/s via the ethernet port. 
		
\section{Position Control}


The positioning of the antenna to the desired source is achieved with the control of two three-phase induction motors \citep{gawronski2008modeling}.
The antenna mount is altazimuthal and this provides independence on each of the axes for the positioning.


The tools used in the development of the position control system are CMake, Ceedling, and PYTHON. PYTHON is used to generate trajectories, astronomical calculations, and the development of the application logic. Ceedling is a framework used for software mocks, unit testing, and code coverage reporting. CMake performs the compilation of embedded systems independently. 

\begin{figure}
    \centering	\includegraphics[width=0.8\columnwidth,keepaspectratio]{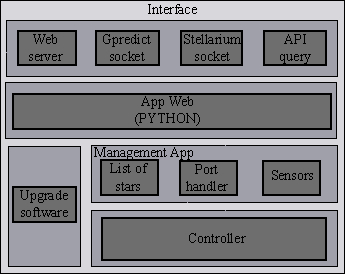}
    \caption{Layered model of the position controller.}
    \label{fig.arqitectura}
\end{figure}

The controller software is realised in a layered architecture detailed in the Figure~\ref{fig.arqitectura}.
There, the interface layer is in charged of bidirectional communications with the outside world. 
The web app manages the connections and communication with the antenna controller hardware. In addition, it is responsible for updating radio sources, conflict resolution and user administration. Furthermore, it manages the observation schedule, the system telemetry data collection, the calibration noise source and sends the position data to the controller. This application is implemented on an embedded Linux system (beaglebone black-Cortex A9). 
The interface layer is currently under development.

On the other hand, the controller layer is responsible for obtaining the environmental parameters and position control. 
This layer is composed by a distributed system made up of three RP2040 microcontrollers. Two of them are intended for position control and the third for environmental telemetry. 
In the case of the microcontrollers used in the control stage, one core is dedicated to the use of I2C (Inter-Integrated Circuit) communication, while the other is in charge of control.

The controller layer has a bidirectional communication with the management app layer. In addition, the controller layer sends the telemetry data of the environment and of the control to the management app layer, and receives from the latter the position to be pointed, together with the positioning events. These events can be to activate/deactivate the Wait, tracking or Alarm state. In Wait or Alarm mode the positioning is at the zenith. 
The angular control system is the same for both angles (azimuth and height).

\begin{figure}
    \centering	\includegraphics[width=0.8\columnwidth,keepaspectratio]{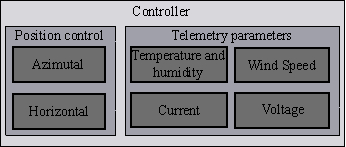}
    \caption{Block diagram of the controller layer.}
    \label{fig.controller_app}
\end{figure}



Communication between the cores occurs through the use of a queue. Each time the core receives a command via I2C, it performs a validation of the angular command, and if this command is valid, it is sent to the second core.

The microcontroller has two more cores, called PIOs (Programmable Input Output). These PIO machines, in turn, have 4 slots of 32 memory instructions, where each of them can execute instructions in parallel to the main cores of the application. These PIO machines communicate with the processor via IRQ (Interruption Request), or by clearing the buffers. In addition, they can communicate with GPIO (General Purpose Input Output). 

The initialisation software consists of a fuzzy logic self-tuning PI controller \citep{Kanagaraj}.

The reading of the environmental and electrical variables is performed every two seconds by the main core, by the architecture known as superloop. 
For the humidity reading the DHT11 sensor was used. On the other hand, the code coverage was carried out by Ceedling, obtaining the expected results.


From the interface layer, the Gpredict \footnote{http://gpredict.oz9aec.net/} and Stellarium\footnote{https://stellarium.org/es/} software stages have been developed, with the API (Application Programming Interface) and the web server still to be defined. The interface layer receives coordinates and transforms them to the local system adopted by the antenna position. In the case of Stellarium, these are given in right ascension and declination. 
To carry out the coordinate transformation, a routine has been created in the C programming language. This routine was tested on the ATMEGA328P microcontroller (with an 8-bit architecture, see algorithm in \citeauthor{mechanorbital}). 
The error of the sidereal time calculation was approximately 70 msec.

In the case of Gpredict, the coordinates come in local altazimuthal coordinates and do not require any transformation to local coordinates.

\section{Conclusions}

In conclusion, the state of progress of the MIA interferometer is promising. Acceptable preliminary results were obtained in each stage. The challenge now is to meet the objectives set for this first stage. The second stage will be to expand the number of antennas of the interferometer. The scientific potential that MIA will offer is promising, and the expectations of the scientific community are high.


\end{document}